\documentclass{article}
\usepackage{spconf,amsmath,graphicx}
\usepackage{booktabs}
\usepackage{siunitx}
\usepackage{graphicx}
\usepackage{bm}
\usepackage{subcaption} 
\usepackage[sort]{cite}
\usepackage[hidelinks]{hyperref}
\usepackage{flushend}

\title{COMPARATIVE ANALYSIS OF DISCRIMINATIVE DEEP LEARNING-BASED NOISE REDUCTION METHODS IN LOW SNR SCENARIOS}
\name{Shrishti Saha Shetu, Emanu\"{e}l A. P. Habets and Andreas Brendel}
\address{Fraunhofer IIS, Am Wolfsmantel 33, 91058 Erlangen, Germany \\
\small \textit{\{shrishti.saha.shetu, emanuel.habets, andreas.brendel\}@iis.fraunhofer.de}}
\begin{document}
%
\maketitle
\begin{abstract}
In this study, we conduct a comparative analysis of deep learning-based noise reduction methods in low signal-to-noise ratio (SNR) scenarios. Our investigation primarily focuses on five key aspects: The impact of training data, the influence of various loss functions, the effectiveness of direct and indirect speech estimation techniques, the efficacy of masking, mapping, and deep filtering methodologies, and the exploration of different model capacities on noise reduction performance and speech quality. Through comprehensive experimentation, we provide insights into the strengths, weaknesses, and applicability of these methods in low SNR environments. The findings derived from our analysis are intended to assist both researchers and practitioners in selecting better techniques tailored to their specific applications within the domain of low SNR noise reduction.
\end{abstract}
\vspace{-.2cm}
\section{Introduction}
\label{sec:intro}
\vspace{-.2cm}
Recent advancements in deep learning-based speech enhancement (SE) methods can effectively improve speech quality by adeptly suppressing noise, thereby providing significant benefits across various applications. State-of-the-art (SOTA) methods primarily employ discriminative approaches, with a focus on enhancing signal quality in moderate signal-to-noise ratio (SNR) scenarios \cite{liu2023mask,schroter2022deepfilternet2, choi2021real,zhao2022frcrn,shetu2023ultra}. However, in low SNR conditions (e.g., below \mbox{-5}~dB ), the desired speech signal is often entirely masked by noise, posing a challenge for existing methods to improve perceptual speech quality while effectively suppressing noise \cite{hao2020masking}. Although constantly low SNR levels are less common in everyday voice communication or human-machine interaction scenarios, segmental SNRs within this range are to be expected in real-world recordings, and low SNR situations are also prevalent in various industrial applications, such as in construction, healthcare facilities, and industrial settings.

In recent years, numerous methods have been proposed to enhance the performance of DNN-based SE algorithms for low SNR scenarios \cite{hao2020masking,gao16_interspeech}. Additionally, several general studies on DNN-based SE methods have been published recently, focusing on aspects such as loss functions, training targets, and various model architectures \cite{braun2021consolidated,kolbaek2020loss,6887314,nossier2020mapping,zheng2023sixty,braun2021towards}.

Despite this, a comprehensive study on low SNR SE, which investigates the most critical aspects that can affect noise reduction performance and speech quality improvements, is notably absent in the literature. Therefore, our study aims to fill this gap by examining different factors that influence the overall SE performance in low SNR conditions. Specifically, we investigate the impact of training data, the influence of various loss functions, the effectiveness of direct and indirect speech estimation techniques, the efficacy of masking, mapping, and deep filtering methodologies, and the exploration of different model capacities.
\vspace{-.2cm}
\section{Methods}
\label{sec:Methods}
\vspace{-.2cm}
In our study, we assume an additive signal model in the time domain: $\mathbf{x}(n) = \mathbf{s}(n) + \mathbf{v}(n)$
, where $\mathbf{x}(n)$, $\mathbf{s}(n)$, and $\mathbf{v}(n)$ denote the noisy signal, clean speech, and noise components, respectively. The discrete time index is denoted by $n$, and is omitted for brevity in the remainder of this paper.
We assume a DNN processing framework, which takes $\mathbf{x}$ as the input and estimates the clean speech signal $\mathbf{s}$. We denote the clean speech estimate as $\mathbf{\hat{s}}$.
\vspace{-.2cm}
\subsection{Learning objectives}
\vspace{-.2cm}
We incorporate four distinct loss functions from existing literature into our learning framework: The scale-invariant signal-to-distortion ratio (SI-SDR) \cite{hu2020dccrn} loss, the multi-scale loss \cite{choi2021real}, the multi-target loss \cite{schroter2022deepfilternet2}, and the joint loss \cite{pandey2020densely}. These loss functions and their variations are widely employed in literature because of their distinctive approaches to emphasizing errors associated with specific signal components. These include addressing time-domain signal distortion, utilizing time-frequency (TF)-domain distance metrics, and applying frame-wise time-domain or TF-domain distance metrics to enhance both the magnitude and phase components of the estimated signal.\\ 

\noindent\textbf{SI-SDR Loss}: The SI-SDR was first proposed in \cite{le2019sdr} to alleviate the weakness of widely used SDR metrics, which evaluates the distortion of the enhanced speech signal in the time domain, ignoring global attenuation effects. It has also been used as a learning objective in many recent SE methods \cite{hu2020dccrn, luo2019conv}. The SI-SDR loss is defined as:
\begin{align*}
\mathbf{s}_{\text{target}} = \frac{ \hat{\mathbf{s}}^\text{T} \mathbf{s}  }{\|\mathbf{s} \|^2} \mathbf{s}  \hspace{0.9cm}   \mbox{and} &\hspace{0.9cm} \mathbf{e}_{\text{noise}} = \hat{\mathbf{s}} - \mathbf{s}_{\text{target}}
\end{align*}
\begin{equation*} 
    \mathcal{L}_\text{SI-SDR} = 10 \log_{10}\left(\frac{\|\mathbf{s}_{\text{target}} \|^2}{\|\mathbf{e}_{\text{noise}} \|^2}\right).
\end{equation*}

\noindent \textbf{Multi-Scale (MS) Loss}: The MS loss $\mathcal{L}_\text{MS}$ is a combination of an MS loss in time domain $\mathcal{L}_\text{wav}$ based on the cosine similarity (CS) and mean squared error (MSE) in the frequency domain $\mathcal{L}_\text{spec}$, $\mathcal{L}_\text{MS}$= $\mathcal{L}_\text{wav}$ + $\mathcal{L}_\text{spec}$. The time-domain loss can be described as:\vspace{-.2cm}
\begin{equation*}
\vspace{-.1cm}
\mathcal{L}_\text{wav} = \sum_j \frac{1}{K} \sum_{k=1}^{K} \text{CS}\left(\mathbf{s}_{jk}, \hat{\mathbf{s}}_{jk}\right),
\vspace{-.1cm}
\end{equation*}
where $K$ denotes the number of segments, and $j$ is an index, indicating a set of segment lengths in $\{16 \text{ms}, 32 \text{ms},
64 \text{ms},\\ 128\text{ms}\}$ \cite{choi2021real}. The MS loss in the frequency domain is defined as:
\begin{equation*} 
\vspace{-.1cm}
\mathcal{L}_\text{spec} = \sum_i \left\|\left| \mathbf{S}_i \right|^{\alpha} - \left| \widehat{\mathbf{S}_i } \right|^{\alpha}\right\|_\text{F}^2,
\vspace{-.1cm}
\end{equation*}
where $\left\| \text{.}\right\|_\text{F}$ is the Frobenius norm, $\mathbf{S}_i = \text{STFT}_i(\mathbf{s})$ is the $i$-th STFT with window sizes in $\{16 \text{ms}, 32 \text{ms}, 64 \text{ms}\}$ and $\alpha \in(0,1]$ is a power law compression factor, here chosen as 0.3.\\

\noindent\textbf{Multi-Target (MT) Loss}: The MT loss $\mathcal{L}_\text{MT}$ is a modified version of the frequency-domain MS loss. In this loss function, additionally to comparing magnitudes, a phase component is introduced. The MT loss function is defined as:
\begin{equation*}
\vspace{-.1cm}
\mathcal{L}_\text{MT} = \mathcal{L}_\text{spec} + \sum_{i} \left\| \left| \mathbf{S}_i \right|^{\alpha}\odot e^{j\bm{\phi}_\mathbf{S}} - \left| \widehat{\mathbf{S}}_i \right|^{\alpha}\odot e^{j\bm{\phi}_{\widehat{\mathbf{S}}}} \right\|_\text{F}^2,
\vspace{-.1cm}
\end{equation*}
where $\bm{\phi_\mathbf{S}}$ denotes the phase component of $\mathbf{S}$ and $\odot$ denotes the Hadamard product.\\

\noindent \textbf{Joint Loss (JL)}: The joint loss $\mathcal{L}_\text{JL}$ is a combination of the time-domain SI-SDR loss $\mathcal{L}_\text{SI-SDR}$ and the MSE loss between the estimated $\mathbf{\widehat{M}}$ and the oracle complex ratio mask $\mathbf{M}$, i.e. complex ideal ratio mask  \cite{zhao2022frcrn}, as following,
\begin{equation*}
\vspace{-.1cm}
\mathcal{L}_\text{JL} = \mathcal{L}_\text{SI-SDR} +  \text{E}\left[ \left\|\mathbf{M} -  \mathbf{\widehat{M}}\right\|_\text{F}^2 \right],
\vspace{-.2cm}
\end{equation*}
where E$[\cdot]$ represents the expectation operator.
\vspace{-.2cm}
\subsection{DNN Models}
\label{dnn models}
\vspace{-.2cm}
In this study, we evaluate four groups of DNN-based SE methods relying on estimating the time-domain waveform, the magnitude spectrum, the complex-valued spectrum,  or its magnitude and phase separately. In total, we evaluate six different DNN models following various design architectures and model capacities in terms of the number of parameters and multiply-accumulate operations per second (MACS), as specified in Table 1.

 \begin{table}[h]
 \vspace{-.2cm}
\centering
\small
\setlength{\tabcolsep}{3pt} 
\begin{tabular}{l S[table-format=3.0] c S[table-format=2.2] S[table-format=2.2]} \toprule
    {\textbf{Model}} & {\textbf{FFT Length}} & {\textbf{Causality}} & {\textbf{Params (M) }} & {\textbf{GMACS}}\\ \midrule
    CRN            & 512 & \text{Yes} & 17.58 & 2.57 \\
    GCRN           & 512 & \text{Yes} & 9.77 & 2.42  \\ 
    DCCRN          & 512 & No & 3.67 & 11.30  \\
    TaylorSENet    & 320 & \text{Yes} & 5.45 & 6.43 \\ 
    FullSubNet+    & 512 & No& 8.67 & 30.06  \\ 
    DDAEC          & {NA} & No & 4.82 & 18.34  \\ \bottomrule
\end{tabular}
\vspace{-.2cm}
\caption{ \small Specifications of different SOTA DNN models.}
\vspace{-.5cm}
\label{tab:yourlabel}
\end{table}
Magnitude spectrum-based methods estimate a magnitude mask by a DNN model, which is bounded between 0 and 1. The estimated magnitude mask is then multiplied by the magnitude of the noisy speech signal and is finally combined with the noisy phase to estimate the clean speech signal. In our study, we use the CRN model \cite{zheng2023sixty} for this approach. 

In recent years, complex-valued spectrum mapping-based approaches have been quite popular, and many methods \cite{hu2020dccrn,chen2022fullsubnet+} have been proposed to jointly enhance the magnitude and the phase component of the noisy signal. In most of these approaches, either a complex ratio mask (CRM) is estimated, or the real and imaginary part of the complex-valued spectrum of the clean speech signal is directly estimated by a DNN model. In this category, we evaluate three different SOTA models: DCCRN \cite{hu2020dccrn}, GCRN \cite{tan2019learning}, FullSubNet+ \cite{chen2022fullsubnet+}.

In two-stage models, typically, a magnitude mask is estimated in the first stage of the DNN model. The enhanced magnitude spectrum is then used to approximate the clean phase component, either by a second-stage DNN model \cite{schroter2022deepfilternet2,choi2021real} or a model designed-based on classical signal processing methods \cite{song2024phase,4244543}. In our study, we evaluate TaylorSENet \cite{li2022taylor}, which is based on Taylor’s approximation theory. 

From time-domain-based methods, we evaluate the dilated and dense autoencoder (DDAEC) \cite{pandey2020densely}, which is a fully convolutional neural network for real-time speech enhancement in the time domain. 

\begin{figure*}[t]

\centering
\includegraphics[width=.83\textwidth]{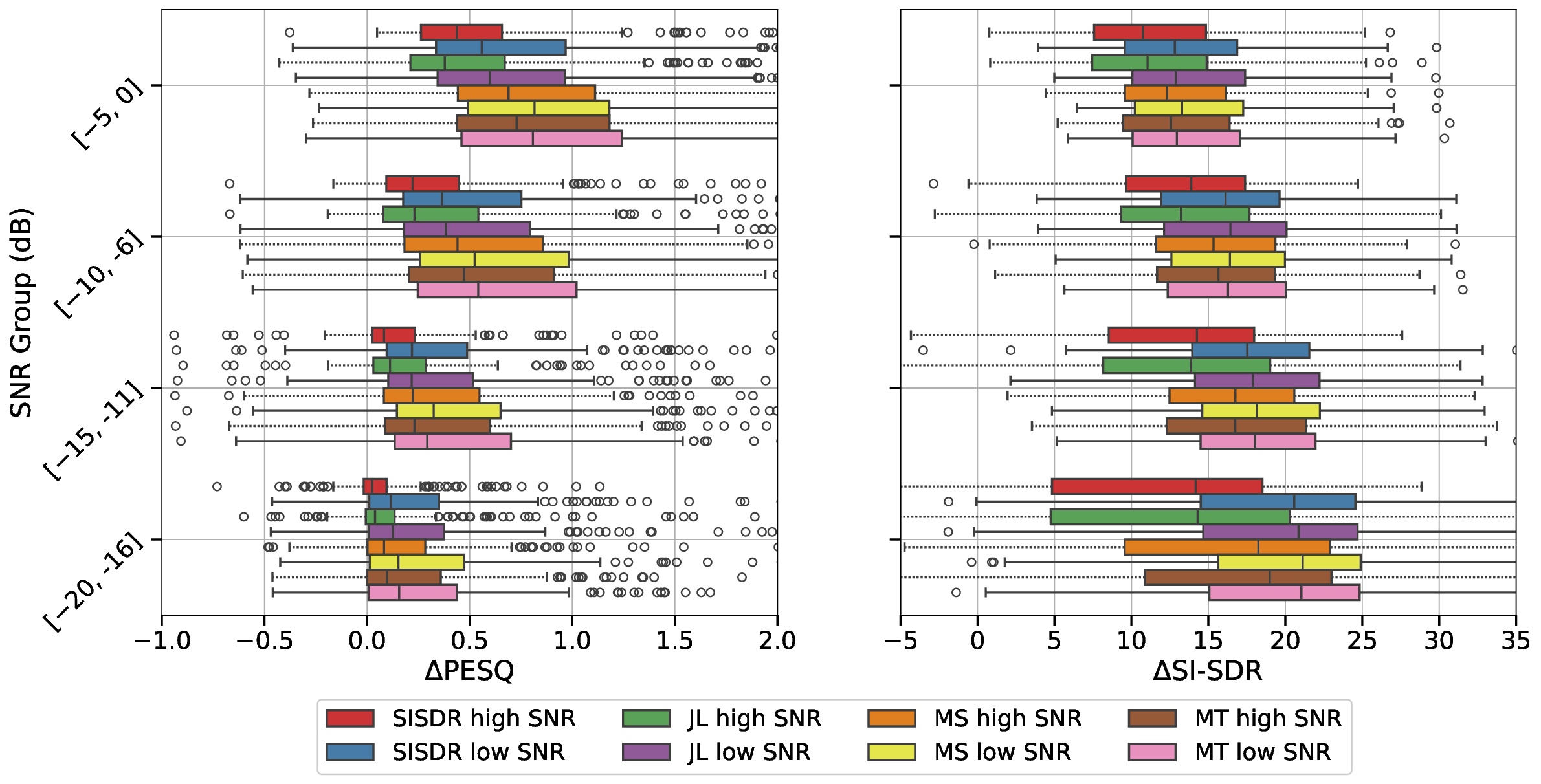}
\vspace{-.2cm}
\caption{ \small PESQ and SI-SDR improvement for the DCCRN model trained on both high (dotted line) and low (solid line) SNR datasets, employing various loss functions. Please note that the legends are labeled according to the model's loss function and training dataset.}
\label{fig:LF}
\vspace{-.5cm}
\end{figure*}

\vspace{-.2cm}
\section{Experiments}
\label{sec:Exp}
\vspace{-.2cm}

\subsection{Implementation Details}
\vspace{-.2cm}
\noindent \textbf{Training Dataset}: We used the Interspeech 2020 DNS Challenge dataset \cite{reddy2020interspeech} to train the DNN models.  We created two separate datasets to focus on high and low SNR scenarios. We created noisy mixtures by randomly selecting and mixing utterances from the clean speech and noise sets at random SNRs: [-25, 0] dB for low SNR and [-5, 30] dB for high SNR. In total, we created a training dataset of around
1000 hrs each for both scenarios. In 50$\%$ of the training dataset, we convolved clean speech with a room impulse response (RIR) randomly selected from the RIR dataset provided in \cite{reddy2020interspeech}.\\

\vspace{-0.2cm}
\noindent \textbf{Evaluation Dataset}: To assess the performance of various models across different configurations, we curated an evaluation dataset comprising 900 samples, each lasting 10~s. As the reference for clean speech, we used the synthetic non-reverb DNS challenge test dataset \cite{reddy2020interspeech}, which contains a total of 150 clean speech samples. Each of these clean speech samples was then mixed randomly with noise at SNR levels ranging from -20 to 0~dB, by using six different noise samples per item randomly selected from the ESC-50 dataset \cite{piczak2015dataset}.  To ensure comprehensive evaluation, we categorized this dataset into four SNR groups: Group 1 [-5, 0], Group 2 [-10, -6], Group 3 [-15, -11], and Group 4 [-20, -16] dB. \\

\vspace{-0.2cm}
\noindent \textbf{Training Targets and Parameters}: In our study, we always used the clean speech signal $\mathbf{s}$ as our training target. We either directly or indirectly estimated the clean speech signal $\mathbf{\hat{s}}$ by using CRM \cite{williamson2016complex}, mapping \cite{nossier2020mapping} or deep filtering (DF) \cite{mack2019deep} approaches. For indirect speech estimation, we first estimated the noise components $\mathbf{\hat{v}}$, which are then subtracted from the noisy signal $\mathbf{x}$ to estimate the clean speech signal $\mathbf{\hat{s}}$.

In our evaluation, we used PyTorch to train the models with the Adam optimizer. The initial learning rate was set to 0.002 and was halved when the validation loss went up within two epochs. The window lengths and hop sizes were chosen as suggested in the original works. \\

\vspace{-0.2cm}
\noindent \textbf{Evaluation Metrics}: To evaluate different methods, we used SI-SDR \cite{le2019sdr} and PESQ \cite{rix2001perceptual} as the main objective metrics. Previous studies have indicated that the PESQ correlates more with perceived speech quality, while SI-SDR correlates more with background noise suppression \cite{shetu2023ultra,torcoli2021objective}. Other metrics, such as STOI \cite{taal2010short} and segmental SNR, followed similar trends as SI-SDR and PESQ and are, hence, omitted in the presentation of the experimental results.

\begin{figure}[t]
\centering
\includegraphics[width=.43\textwidth]{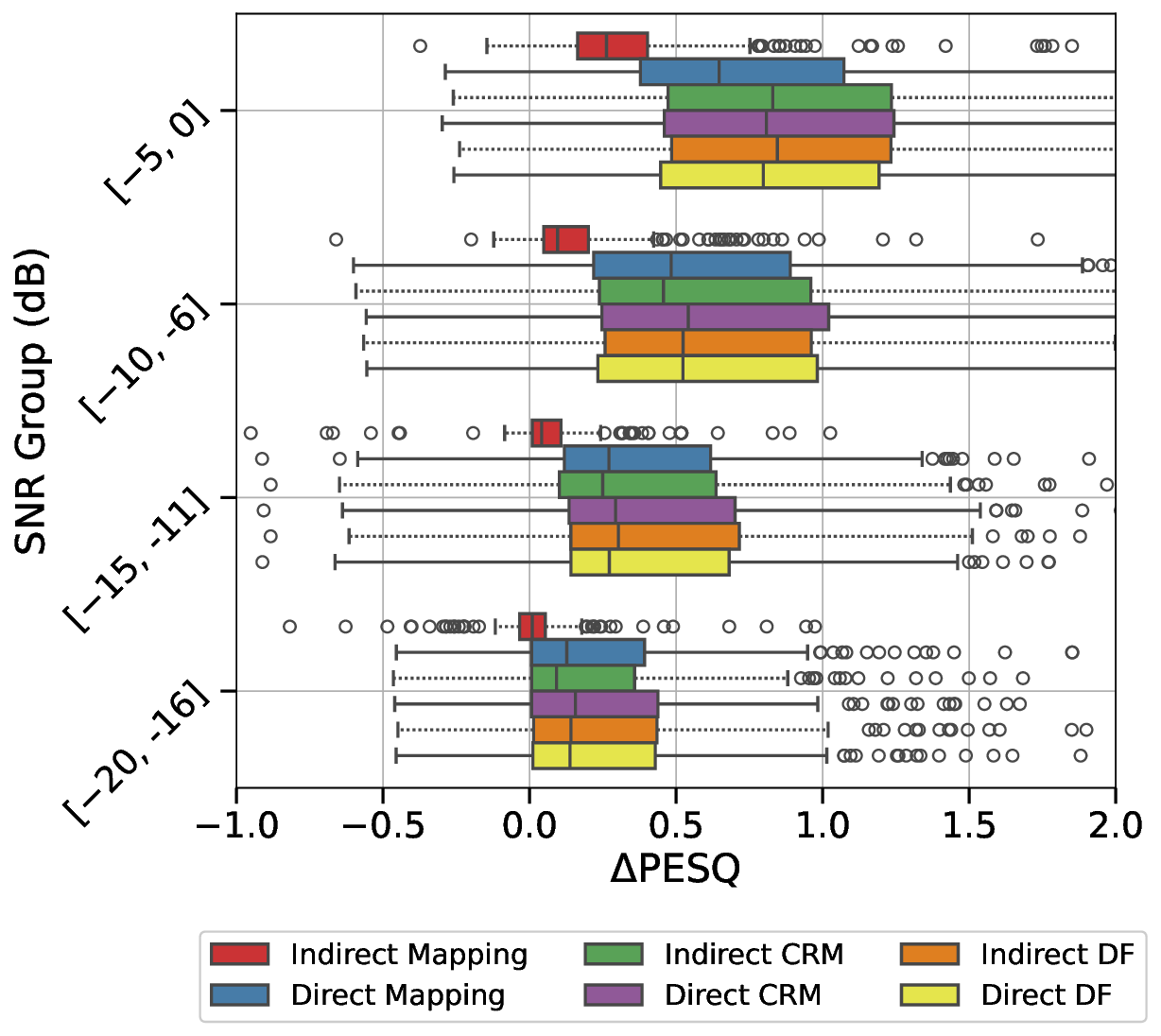}
\vspace{-.2cm}
\caption{\small PESQ improvement for the DCCRN model trained with the low SNR training dataset for indirect (dotted line) and direct (solid line) speech estimation with mapping, CRM and DF approaches.}
\label{fig:TT}
\vspace{-.6cm}
\end{figure}

\begin{figure}[t]
\centering
\includegraphics[width=.4\textwidth]{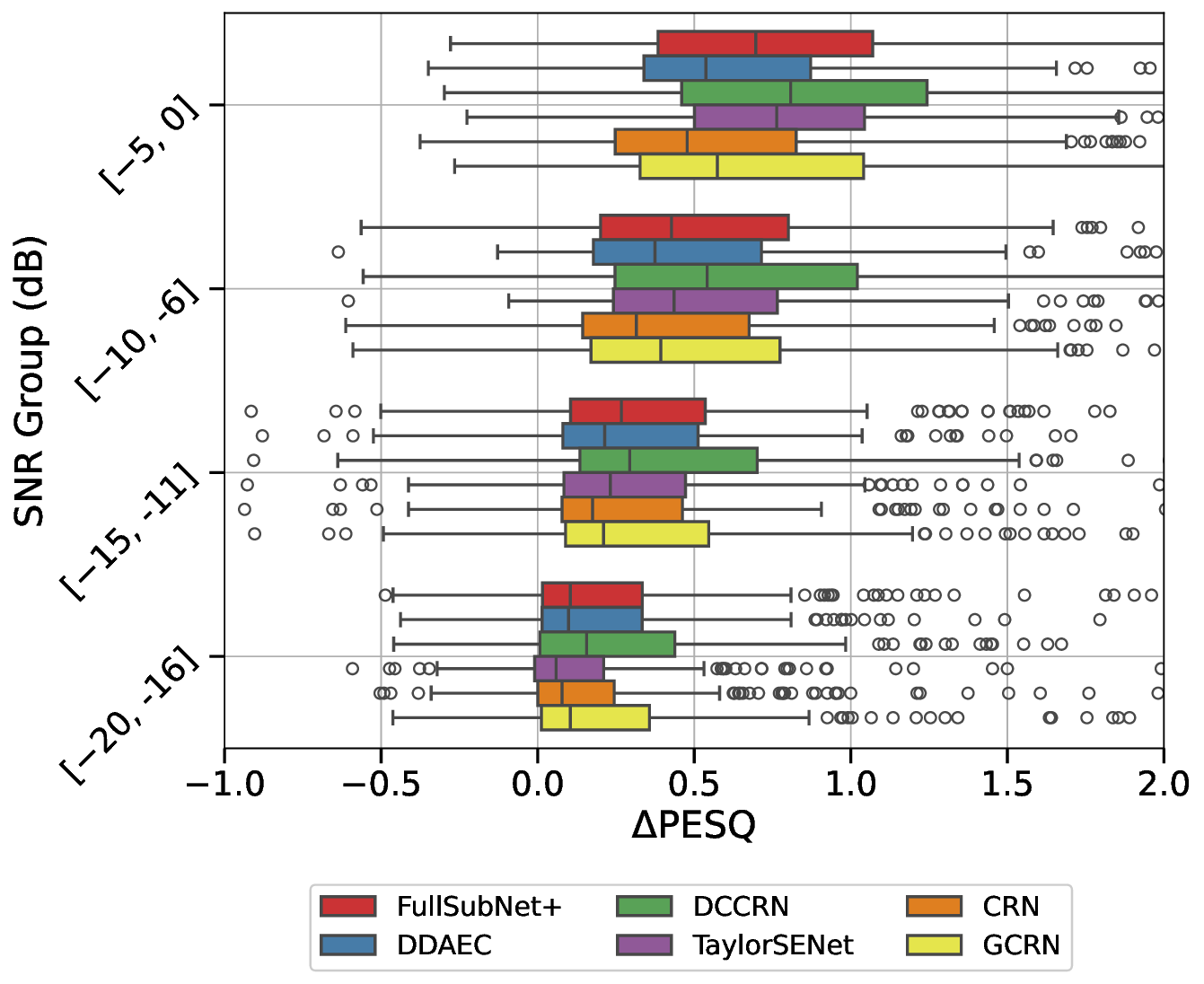}
\vspace{-.2cm}
\caption{\small PESQ improvement for different SOTA models trained with the low SNR training dataset for direct speech estimation using CRM masking method (models are ordered in terms of ascending MACS).}
\label{fig:MC}
\vspace{-.5cm}
\end{figure}

\vspace{-.3cm}
\subsection{Experimental Design, Results and Discussion}
\label{sec:RD}
\vspace{-.1cm}

\noindent \textbf{(a) Impact of Loss Functions and Training Dataset}: Here, we separately trained the DCCRN model with the low and high SNR datasets and together with all four loss functions, resulting in 8 different models (4 for each low and high SNR training dataset). For this experiment, we directly estimated speech using CRM masking as described in \cite{hu2020dccrn}.

The results in Fig.~\ref{fig:LF} show that the training dataset has a significant impact on the overall SE performance. Models trained with low SNR data outperform those trained with high SNR data in all SNR groups, improving both PESQ and SI-SDR. In terms of loss functions, the MS and MT loss functions clearly outperform the joint loss (JL) and SI-SDR loss functions.\\ 

\noindent \textbf{(b) Impact of Training Targets}: As previously described, we used direct and indirect speech estimation to obtain the clean speech estimate $\mathbf{\hat{s}}$. We trained the DCCRN model separately with both of these techniques in combination with the CRM, mapping, and DF approaches and using only the low SNR dataset, resulting in 6 different models.

We can see in Fig.~\ref{fig:TT} that direct speech estimation always outperforms indirect speech estimation, except for the models trained with DF approaches. For mapping-based methods, we observe that the indirect speech mapping approaches clearly lag behind in all SNR groups. This indicates that, in general, indirect speech estimation by noise mapping is a much more difficult learning task than direct speech mapping.  
The DF and CRM masking methods perform almost similarly for direct speech estimation. However, for indirect speech estimation, the DF approach outperforms the masking method. This indicates that in low SNR scenarios, indirect speech estimation by noise estimation could be beneficial, as suggested in \cite{liu2023mask}, if we can facilitate the DNN model to learn reasonable temporal and spectral context information. \\

\noindent \textbf{(c) Impact of Model Capacities}: In the literature, it has been shown that different model designs, number of parameters, and amount of temporal context have an impact on the overall SE performance \cite{zheng2023sixty}. To evaluate these aspects, we trained all 6 different models described in Section \ref{dnn models} only with the low SNR dataset and their default configurations.

In the results of Fig.~\ref{fig:MC}, we can observe that the DCCRN model clearly outperforms all other approaches in all SNR groups. This result might be explained by our observation that reasonable temporal and spectral context information is needed for high-quality speech enhancement at low SNR. The DCCRN model is a non-causal model with a lookahead of 40 ms, which is the highest among all the considered models, with a moderate complexity of 11.30 GMACS. All the causal and low-complexity models perform reasonably well for higher SNRs but perform poorly at very low SNRs (below -10~dB).

\begin{figure}[t]
    \centering
    \begin{subfigure}{0.44\linewidth}
        \includegraphics[width=\linewidth]{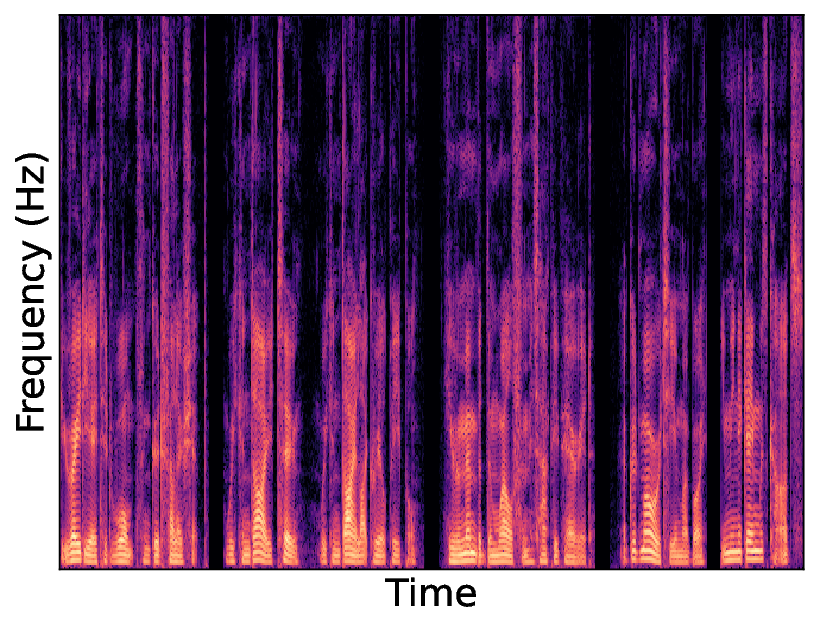}
        \caption{\small Clean}
        \label{fig:sub1}
    \end{subfigure}
    \begin{subfigure}{0.44\linewidth}
        \includegraphics[width=\linewidth]{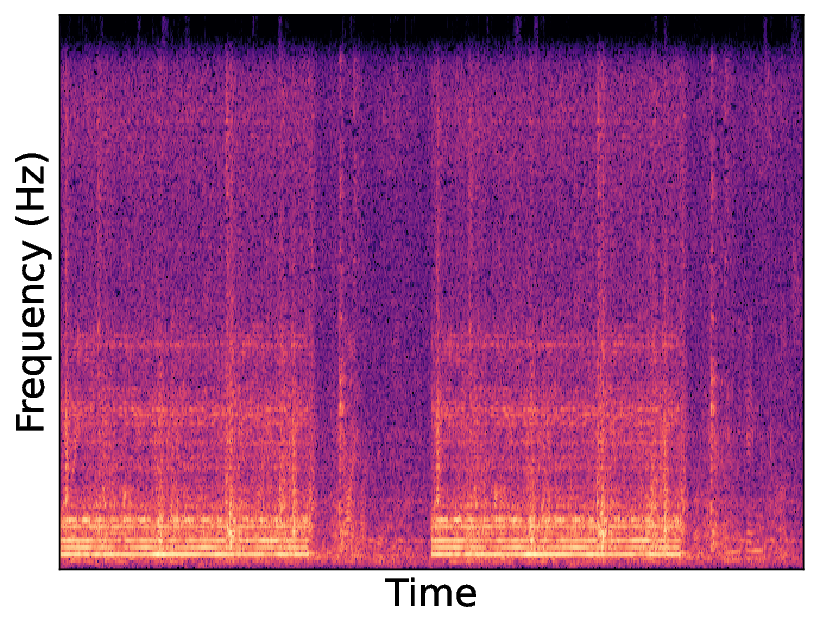}
        \caption{\small Noisy}
        \label{fig:sub2}
    \end{subfigure}
    \\
    \begin{subfigure}{0.44\linewidth}
        \includegraphics[width=\linewidth]{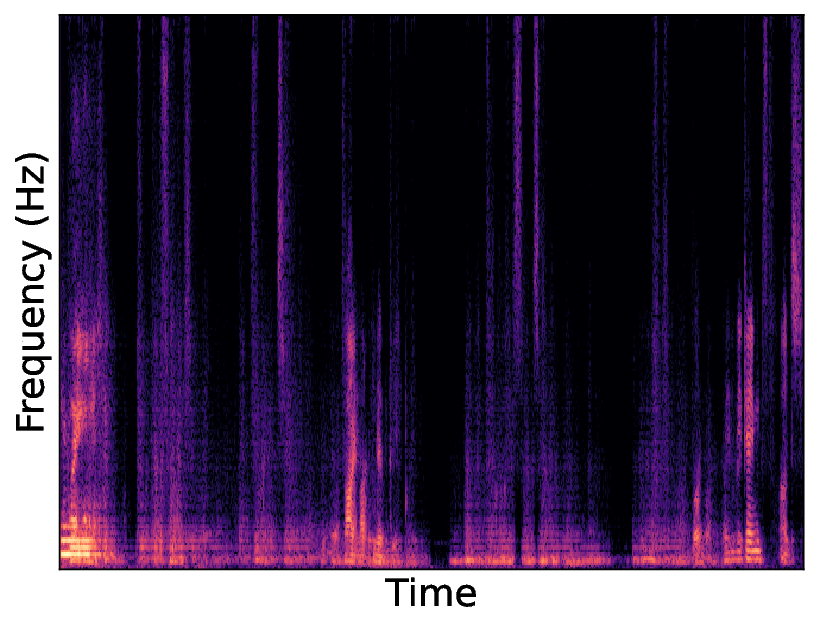}
        \caption{\small DCCRN Model}
        \label{fig:sub3}
    \end{subfigure}
    \begin{subfigure}{0.44\linewidth}
        \includegraphics[width=\linewidth]{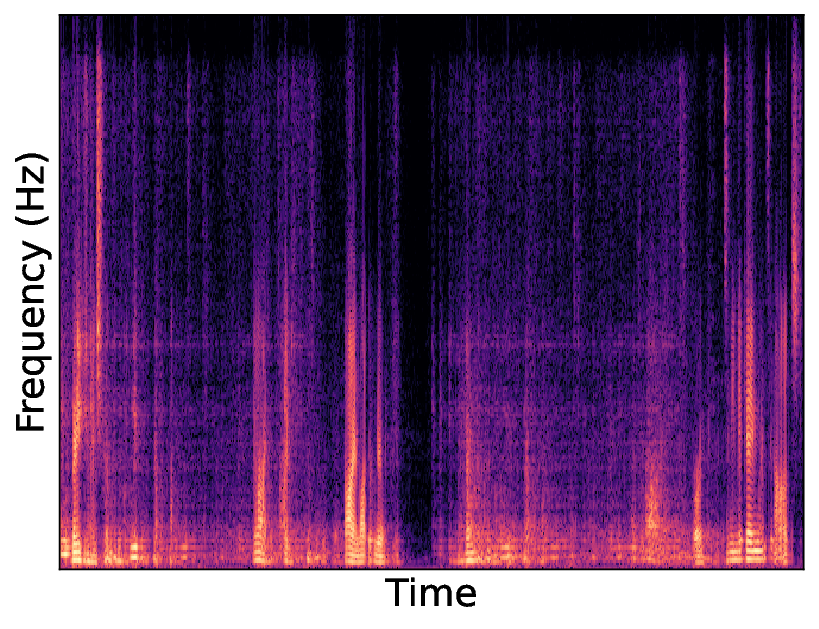}
        \caption{\small  DDAEC Model}
        \label{fig:sub4}
    \end{subfigure}
    \vspace{-.2cm}
    \caption{ \small Example of a (a) clean speech signal (b) masked by strong noise at -14~dB SNR and (c,d) estimated clean speech signal.}
    \label{fig:subplot}
    \vspace{-.4cm}
\end{figure}  

\noindent \textbf{Discussion}: A key takeaway from these experiments is that all studied SOTA methods at various configurations perform quite well for high SNR scenarios. The median PESQ improvement for the SNR range [0, -10]~dB (SNR Groups 1 and 2) is more than 0.7. However, the $\Delta \text{PESQ}$ drops significantly for SNRs below -10~dB. For SI-SDR improvement, we see a reverse trend: The $\Delta \text{SI-SDR}$ increases for decreasing input SNR. However, globally, the SI-SDR values for the enhanced signal decreases as well, similar to $\Delta \text{PESQ}$. 
These results align with our hypothesis that PESQ is more sensitive toward speech quality, whereas SI-SDR tends to provide an objective assessment that may not always encompass all aspects of speech quality.

Our informal subjective listening also aligns with the objective results in high SNR scenarios (for SNR $>$ -5~dB), as all studied methods can enhance the perceptual speech quality significantly in this SNR range. However, for SNRs below \mbox{-5}~dB, the objective results do not always reflect the subjective quality: In many scenarios, where the speech is completely masked by strong noise, as shown in Fig.~\ref{fig:subplot}, none of the studied SOTA methods can enhance the overall speech quality. However, the SI-SDR and PESQ improvements still show significant values. Some listening samples can be found here: \url{https://fhgainr.github.io/lowsnrstudy/}.

\vspace{-.4cm}
\section{Conclusion}
\label{sec:Con}
\vspace{-.2cm}
In this study, we highlight the limitations of SOTA  discriminative methods in very low SNR scenarios. While designing models with high temporal and spectral context and training them with low SNR datasets can enhance SE performance to some degree, improving perceptual speech quality in cases where speech is totally masked by noise remains a challenge for SOTA discriminative approaches. We suggest that future research should explore generative methods to address these challenges effectively in very low SNR scenarios. 

\let\oldbibliography\thebibliography
\renewcommand{\thebibliography}[1]{%
  \oldbibliography{#1}%
  \footnotesize
  \setlength{\itemsep}{0pt}%
}
\bibliographystyle{IEEEbib}
\bibliography{strings,refs}

\end{document}